\documentclass[a4paper,11pt]{article}
\usepackage{jinstpub}
\usepackage{siunitx}

\title{Competition between Increasing and Decreasing Effects of the Afterpulsing Rate of PMTs during Night-Sky Observations}

\author[a]{Takuto Kiyomoto}
\author[a]{Tsutomu Nagayoshi}
\author[b]{Shunsuke Sakurai}
\author[c]{Mitsunari Takahashi%
\footnote{Corresponding author.}}
\author[d]{Tokonatsu Yamamoto}
\author[e]{Alice Donini}
\author[b]{Yusuke Inome}
\author[b]{Yukiho Kobayashi}
\author[b]{Daniel Mazin}
\author[f]{Razmik Mirzoyan}
\author[f]{Seiya Nozaki}
\author[b]{Hideyuki Ohoka}
\author[c,g]{Akira Okumura}
\author[b]{Takayuki Saito}
\author[b]{Ryuji Takeishi}
\author[b,f]{Masahiro Teshima}
\author{for the CTAO LST Project}

\affiliation[a]{Grad. Sch. of Sci. and Eng., Saitama University,\\
255 Simo-Ohkubo, Sakura-ku, Saitama, 338-8570, Japan}
\affiliation[b]{Institute for Cosmic Ray Research, University of Tokyo,\\
5-1-5, Kashiwa-no-ha, Kashiwa, Chiba 277-8582, Japan}
\affiliation[c]{Institute for Space--Earth Environmental Research, Nagoya University,\\
Furo-cho, Chikusa-ku, Nagoya 464-8601, Japan}
\affiliation[d]{Department of Physics, Konan University,\\
Kobe, Hyogo, 658-8501, Japan}
\affiliation[e]{Istituto Nazionale di Astrofisica -- Osservatorio Astronomico di Roma,\\
Via di Frascati 33, 00078, Monteporzio Catone, Italy}
\affiliation[f]{Max-Planck-Institut f\"ur Physik,\\
Boltzmannstr. 8, 85748 Garching, Germany}
\affiliation[g]{Kobayashi--Maskawa Institute for the Origin of Particles and the Universe, Nagoya University\\Furo-cho, Chikusa-ku, Nagoya 464-8602, Japan}

\emailAdd{mitsunari.takahashi@cta-consortium.org}

\abstract{Photomultiplier tubes (PMTs) have been widely used in imaging atmospheric Cherenkov telescopes (IACTs). The Large-Sized Telescopes (LSTs) of the Cherenkov Telescope Array Observatory (CTAO), the latest-generation IACTs, are optimized for challenging observations of low-energy gamma rays, specifically in the 20 to 150 GeV range. To this end, PMTs with an exceptionally low afterpulsing rate have been developed and installed. However, the afterpulsing rate increases over time due to the infiltration of atmospheric molecules, particularly helium, into the tube. Interestingly, we found that the afterpulsing rate decreases when PMTs are operated at high voltage and exposed to light---a condition naturally met during IACT observations. To evaluate the latest instrument response, after five years of operation, we removed several PMTs from the first LST, which is currently the only operational telescope among the CTAO instruments. Our laboratory measurements showed no increase in afterpulsing compared to pre-installation values. This suggests that the decrease in afterpulsing during operation offsets the increase, thereby maintaining the long-term performance of the PMTs.}

\keywords{Cherenkov detectors, Gamma telescopes, Photon detectors for UV, visible and IR photons (vacuum)}

\begin{document}
\maketitle
\flushbottom

\section{Introduction}
\label{sec:intro}
PMTs are crucial in photon-counting experiments because of their high detection efficiency and low noise levels. They have been successfully applied in IACTs, which observe very-high-energy ($\gtrsim \SI{100}{GeV}$) gamma and cosmic rays from the universe.
When such a particle (referred to as the primary particle) enters the atmosphere, it interacts with atmospheric nuclei and produces a cascade of secondary particles, commonly referred to as an air shower. It radiates Cherenkov light, mainly in the ultraviolet--visible waveband. The Cherenkov photons, arriving on the ground with a typical duration of $\sim \SI{3}{ns}$, are collected through optical mirrors and subsequently detected by a camera on the focal plane. The camera comprises multiple photodetector pixels to record an image that shows the direction from which the photons reach the telescope. The image is then used to reconstruct the nature of the primary particle. 
It is worth noting that the number of Cherenkov photons is roughly proportional to the energy of the primary gamma ray.
This feature enables energy estimation but makes observing an air shower from relatively low-energy gamma rays difficult due to the low surface density of Cherenkov light, for example $O(10)\,\rm{photons/m^2}$, for a 100-GeV gamma ray~\cite{BERNLOHR2000255}. Therefore, IACTs require photodetectors with both high sensitivity and low noise. PMTs are well-suited to meet this requirement.
\paragraph{}
PMTs are employed in LSTs and Medium-Sized Telescopes of the CTAO~\cite{WhitePaper}, the latest generation IACTs for gamma-ray observation currently under construction\footnote{https://www.ctao.org}. At present, only the first LST (LST-1) is in operation among the CTAO telescopes.
The LSTs are optimized for relatively low-energy observations, particularly in the range of 20 GeV to 150 GeV. This requires PMTs with extremely low false-signal pulsing rates, in addition to a large mirror dish with a diameter of 23 m.

Afterpulsing in PMTs, caused by gas molecules ionized by accelerated electrons, generates false signals through ion feedback~\cite{PMThandbook}. In the case of LSTs, this process is mainly triggered by night sky photons, which are dominated by airglow and zodiacal light~\cite{NSB_LaPalma}, with a rate of $\sim \SI{250}{MHz/pixel}$.
Hamamatsu Photonics K.K. (HPK)\footnote{https://www.hamamatsu.com} and the authors of~\cite{MIRZOYAN2016640,MIRZOYAN2017603} have developed novel PMTs (R11920-100 for LST-1 and R12992-100 for subsequent telescopes) with an exceptionally low afterpulsing rate, at less than $2\times 10^{-4}$~\cite{TOYAMA2015280,MIRZOYAN2016640,MIRZOYAN2017603,2021NIMPA100765413T}. In this paper, the afterpulsing {rate is defined as the afterpulsing count per photoelectron input, excluding afterpulses with a charge smaller than 4 photoelectron equivalents\footnote{For the amplitude below 4 photoelectrons, the fake signals are dominated by accidentally coincident night sky photons and the afterpulses can be ignored.}. This performance was also confirmed through quality control of the LST-1 PMTs~\cite{PMTModule}.

However, the afterpulsing rate increases with time as a result of atmospheric molecules, particularly helium, penetrating the tube~\cite{Coates_1973,Bartlett_1981}. This increase, which has also been observed in R11920-100 and R12992-100~\cite{PMTModule}, could degrade the energy threshold of the LSTs during their 20-year operational lifetime. Interestingly, we found that the afterpulsing rate decreases when PMTs are operated at high voltage with light exposure. Contrary to experiments conducted in closed environments, this condition naturally met during IACT observations due to the night sky photons. This decrease can be explained by the ionization of residual molecules inside the tube through the ion-feedback process itself. Some of the ionized molecules are captured by the photocathode and consequently removed from the vacuum.} Given these two opposing effects, it is important to determine the latest rate of the LST-1 PMTs to evaluate the telescope's performance.

\section{Measurement}
To evaluate the afterpulsing rate in a controlled environment, we removed several PMTs from LST-1 in October 2023. These tubes were installed shortly before or after the telescope’s first light in December 2018 and subsequently underwent five years of observation\footnote{The telescope is not operated under daylight, strong moonlight, and clouds.}. We located the PMTs in a dark box and illuminated them by fast light pulses with the full width at half maximum of 800--920 ps~\cite{InomePulser} corresponding to roughly 50 photoelectrons. The high voltage was adjusted to achieve a PMT gain of 40000. Then, we counted pulses in the output waveform within 0--3 \textmu s after each light pulse. More details of the setup can be found in~\cite{PMTModule}.
In addition to these removed PMTs (hereafter ``Used'' PMTs), we also measured some ``Spare'' PMTs that had barely been used since their production in 2013. By comparing the results of these two samples, we evaluated the self-cleaning effect that occurs during regular operation.

\section{Results}
In the present paper, we report the results of the measurement of ten Used PMTs and twelve Spare PMTs, which have been analyzed. For comparison, we include data from the quality control carried out in 2014--2015 and analyzed with the same method.

\subsection{Afterpulsing rate}
\begin{figure}[htbp]
\centering
\includegraphics[width=.88\textwidth]{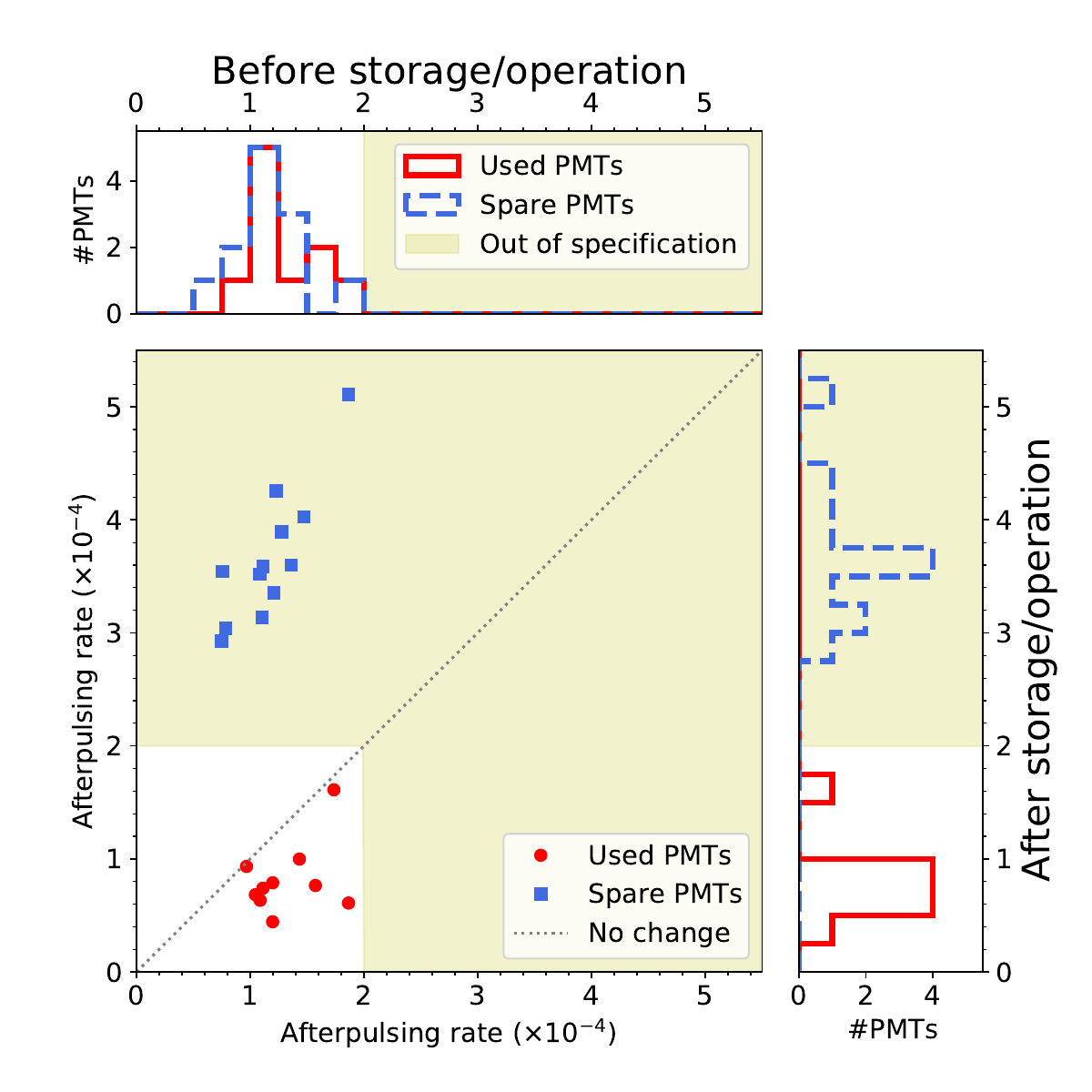}
\caption{\textit{Histograms}: Distribution of the afterpulsing rate integrated for charges $\geq \SI{4}{photoelectrons}$. The same samples of the Spare PMTs (red solid) and the Used ones (blue dashed) appear in both upper and right panels. The range of rates violating the specification, $2 \times 10^{-4}$, is indicated by a yellow hatched region. \textit{Upper panel}: Data taken in 2014--2015, before the long-term storage of the Spare tubes and operation of the Used ones. \textit{Right panel}: Data taken in 2023, after the storage/operation. \textit{Scatter plot}: Afterpulsing rates of the Used PMTs (red circles) and Spare PMTs (blue squares) before the storage/operation compared with those after the storage/operation. The PMT samples and rate values are identical to those in the histograms. The abscissa and ordinate are shared with the upper and right histograms, respectively. The dotted line signifies no change in the rate.\label{fig:AP-distribution}}
\end{figure}

Figure~\ref{fig:AP-distribution} shows the distributions of the afterpulsing rate of the PMT samples. The upper and right panels show the rates measured in 2014--2015 as the quality control and in 2023 after the long-term storage or operation, respectively. In the lower-left panel, the afterpulsing rate of each tube is plotted as a comparison between after and before the interval.

Before storage or operation, both Spare and Used samples exhibited similar afterpulsing rates. After the long interval, while the rate of the Spare samples increased, that of the Used ones slightly decreased and was lower than the specification of the afterpulsing rate $<2 \times 10^{-4}$.

\subsection{Afterpulse arrival time}
\begin{figure}[htbp]
\centering
\includegraphics[width=\textwidth]{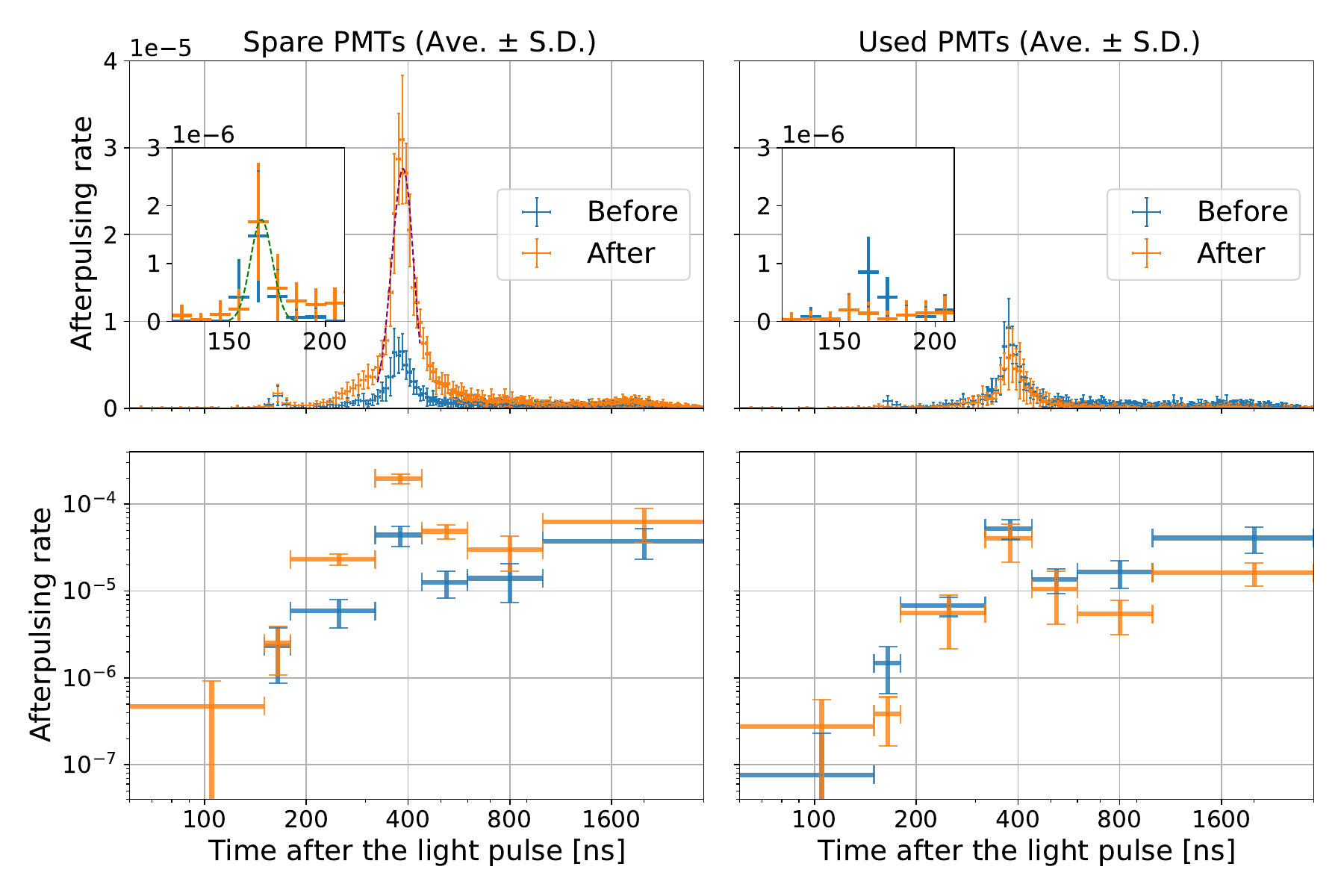}
\caption{Spare (\textit{left panels}) and Used (\textit{right panels}) PMTs' distributions of the afterpulsing arrival time from the light pulse. The afterpulses with a charge $\geq \SI{4}{photoelectrons}$ are counted. The points and error bars in the rate represent the mean and the standard deviation, respectively, of each PMT sample set. The abscissa is in the logarithmic scale. Values are integrated within each bin. \textit{Upper panels}: Time distributions with 5-ns binning on the abscissa and the ordinate in the linear scale. The first peak at 167 ns is shown in an inset. The green and purple dashed curves in the upper-left panel represent the Gaussian fits to the first and second peaks of the data after the storage, respectively. \textit{Lower panels}: Identical data as the upper panels but rebinned to compare the changes within different time domains with the ordinate in the logarithmic scale. \label{fig:AP-time}}
\end{figure}

Figures~\ref{fig:AP-time} presents the distributions of the afterpulse arrival time measured from the light pulse for the Spare PMTs and the Used ones, respectively. The upper and lower panels of each figure display the same data but with different time binnings. 
In both upper panels, two distinct peaks are visible. Fitting the distribution for the Spare PMTs after the storage, which exhibits the largest event number, with a Gaussian function within the ranges of 145--185 ns and 325--435 ns yields peak positions and errors of $166.6 \pm 1.7$ ns and $386.6 \pm 1.7$ ns, respectively.

The afterpulsing rate of the Spare PMTs increased over the nine years since the quality control measurement. Notably, the rate associated with the second peak increased by a factor of $4.4$. In contrast, the Used PMTs do not show any increase in the rate after four years of storage and five years of operation, and it decreased slightly. 

\section{Discussion and Conclusions}
Firstly, we confirmed an increase in afterpulsing for the Spare PMTs. The second peak at 387 ns shows the most noticeable increase. The most plausible candidate of its origin is $\rm{He^+}$, as helium molecules, being small, can permeate the tube~\cite{Coates_1973,Bartlett_1981}, and the ionization cross section of $\rm{He^+}$ is more than 200 times larger than that of $\rm{He^{2+}}$~\cite{Shah_1988}. This peak has long tails that does not follow the Gaussian fit. The reason remains unclear, but these may arise from helium molecules ionized in different spatial regions within the tube~\cite{Coates_1973}.

We next mention possible origins of the first peak in addition. The ions responsible for the first peak at 167 ns must have smaller mass and/or larger charge than $\rm{He^+}$ to arrive at the photocathode earlier. Thus, the candidates are $\rm{H^+}$ and $\rm{H_2^+}$. The possibility of $\rm{He^{2+}}$ is excluded because the first peak increased only by $\sim 8$\% while the second one increased by a factor of 4.4 (See the lower-left panel of Figure~\ref{fig:AP-time}). The afterpulse arrival time $t$ should follow the relation of $t \propto \sqrt{m/q}$ for an ion with mass $m$ and charge $q$ accelerated in a uniform electric field. The expected ratios relative to $\rm{He^+}$ are $t_{\rm{H^+}}/t_{\rm{He^+}}\approx 0.50$ and $t_{\rm{H_2^+}}/t_{\rm{He^+}}\approx 0.71$. The observed ratio of the peak times is $t_1/t_2=\frac{166.6 \pm 1.7}{386.6\pm 1.7}=0.431\pm 0.008$, which is closer to $t_{\rm{H^+}}/t_{\rm{He^+}}$ than $t_{\rm{H_2^+}}/t_{\rm{He^+}}$, but does not exactly match either. This discrepancy may be explained by a non-uniform electric field inside the tube.

Lastly, the afterpulsing of the Used PMTs is clearly lower than the Spare ones and did not show any increase after the long-term storage and operation. This suggests that residual molecules are removed from the vacuum during regular observations and that the increasing and decreasing effects are in equilibrium. The reduction factor is smaller for the second peak, as shown in the right panels of Figure~\ref{fig:AP-time}, than for the ealier and later components. This implies that the helium achieves a different equilibrium state from other molecules due to the infiltration capability. The disappeared molecules are possibly trapped in the photocathode, but their state and fate remain unclear based on our results. To elucidate it further investigation is needed. The level of afterpulsing of the Used samples remains sufficiently low from the perspective of gamma-ray observation and data analysis.

\acknowledgments
We gratefully acknowledge financial support from the following agencies and organisations: KMI, Nagoya University, ICRR, University of Tokyo, JSPS through ``KAKENHI Grant Number 23H04897, 21H04468, 
20KK0067 and 22H01226'', MEXT, Japan; IFNN, Italy; the Spanish Ministry of Science and Innovation and the Spanish Research State Agency through the grant ``FPN: PID2019-107847RB-C41'', the ``CERCA'' program funded by the Generalitat de Catalunya.
We also thank the Instituto de Astrofísica de Canarias (IAC) for the technical support and for the use of the clean room facilities of the IAC, in the central headquarters in La Laguna and of IACTEC.
This research is part of the Project RYC2021-032991-I, funded by MICIN/AEI/10.13039/501100011033, and the European Union ``NextGenerationEU''/RTRP. We also appreciate the information on the PMTs provided by HPK.

\bibliographystyle{JHEP}
\bibliography{biblio.bib}

\providecommand{\href}[2]{#2}\begingroup\raggedright\begin{thebibliography}{10}

\bibitem{BERNLOHR2000255}
K.~Bernlöhr, \emph{{Impact of atmospheric parameters on the atmospheric
  Cherenkov technique}},
  \href{https://doi.org/https://doi.org/10.1016/S0927-6505(99)00093-6}{\emph{Astroparticle
  Physics} {\bfseries 12} (2000) 255}.

\bibitem{WhitePaper}
{The CTA Consortium}, \emph{Science with the Cherenkov Telescope Array}, WORLD
  SCIENTIFIC (Feb., 2018), \href{https://doi.org/10.1142/10986}{10.1142/10986}.

\bibitem{PMThandbook}
E.C.~Hamamatsu Photonics K.~K., \emph{PHOTOMULTIPLIER TUBES --Basics and
  Applications--}, Hamamatsu Photonics K. K., Electron Tube Devision,
  fourth~ed. (2017).

\bibitem{NSB_LaPalma}
C.~Benn and S.~Ellison, \emph{{Brightness of the night sky over La Palma}},
  \href{https://doi.org/https://doi.org/10.1016/S1387-6473(98)00062-1}{\emph{New
  Astronomy Reviews} {\bfseries 42} (1998) 503}.

\bibitem{MIRZOYAN2016640}
R.~Mirzoyan, D.~Müller, Y.~Hanabata, J.~Hose, U.~Menzel, D.~Nakajima et~al.,
  \emph{{Evaluation of Photo Multiplier Tube candidates for the Cherenkov
  Telescope Array}},
  \href{https://doi.org/https://doi.org/10.1016/j.nima.2015.08.030}{\emph{Nuclear
  Instruments and Methods in Physics Research Section A: Accelerators,
  Spectrometers, Detectors and Associated Equipment} {\bfseries 824} (2016)
  640}.

\bibitem{MIRZOYAN2017603}
R.~Mirzoyan, D.~Müller, J.~Hose, U.~Menzel, D.~Nakajima, M.~Takahashi et~al.,
  \emph{{Evaluation of novel PMTs of worldwide best parameters for the CTA
  project}},
  \href{https://doi.org/https://doi.org/10.1016/j.nima.2016.06.080}{\emph{Nuclear
  Instruments and Methods in Physics Research Section A: Accelerators,
  Spectrometers, Detectors and Associated Equipment} {\bfseries 845} (2017)
  603}.

\bibitem{TOYAMA2015280}
T.~Toyama, Y.~Hanabata, J.~Hose, U.~Menzel, R.~Mirzoyan, D.~Nakajima et~al.,
  \emph{Evaluation of the basic propertied of the novel 1.5in. size {PMT}s from
  {Hamamatsu Photonics and Electron Tubes Enterprises}},
  \href{https://doi.org/https://doi.org/10.1016/j.nima.2014.12.070}{\emph{Nuclear
  Instruments and Methods in Physics Research Section A: Accelerators,
  Spectrometers, Detectors and Associated Equipment} {\bfseries 787} (2015)
  280}.

\bibitem{2021NIMPA100765413T}
A.~{Tsiahina}, P.~{Jean}, J.F.~{Olive}, J.~{Kn\"{o}dlseder}, C.~{Marty},
  T.~{Ravel} et~al., \emph{{Measurement of performance of the NectarCAM
  photodetectors}},
  \href{https://doi.org/10.1016/j.nima.2021.165413}{\emph{Nuclear Instruments
  and Methods in Physics Research A} {\bfseries 1007} (2021) 165413}
  [\href{https://arxiv.org/abs/2110.06030}{{\ttfamily 2110.06030}}].

\bibitem{PMTModule}
T.~Saito, M.~Takahashi, Y.~Inome, H.~Abe, M.~Artero, O.~Blanch et~al.,
  \emph{{Development and quality control of PMT modules for the large-sized
  telescopes of the Cherenkov Telescope Array Observatory}},
  \href{https://doi.org/https://doi.org/10.1016/j.nima.2025.170229}{\emph{Nuclear
  Instruments and Methods in Physics Research Section A: Accelerators,
  Spectrometers, Detectors and Associated Equipment} {\bfseries 1073} (2025)
  170229}.

\bibitem{Coates_1973}
P.B.~Coates, \emph{The origins of afterpulses in photomultipliers},
  \href{https://doi.org/10.1088/0022-3727/6/10/301}{\emph{Journal of Physics D:
  Applied Physics} {\bfseries 6} (1973) 1159}.

\bibitem{Bartlett_1981}
D.F.~Bartlett, A.L.~Duncan and J.R.~Elliott, \emph{Afterpulses in a
  photomultiplier tube poisoned with helium},
  \href{https://doi.org/10.1063/1.1136585}{\emph{Review of Scientific
  Instruments} {\bfseries 52} (1981) 265}
  [\href{https://arxiv.org/abs/https://pubs.aip.org/aip/rsi/article-pdf/52/2/265/19306319/265\_1\_online.pdf}{{\ttfamily
  https://pubs.aip.org/aip/rsi/article-pdf/52/2/265/19306319/265\_1\_online.pdf}}].

\bibitem{InomePulser}
Y.~Inome, T.~Yamamoto, M.~Teshima, H.~Ohoka, D.~Nakajima and R.~Mirzoyan,
  \emph{Development of a hundred-picoseconds pulse laser as a calibration
  source},  in \emph{2017 IEEE Nuclear Science Symposium and Medical Imaging
  Conference (NSS/MIC)}, pp.~1--2, 2017,
  \href{https://doi.org/10.1109/NSSMIC.2017.8533096}{DOI}.

\bibitem{Shah_1988}
M.B.~Shah, D.S.~Elliott, P.~McCallion and H.B.~Gilbody, \emph{Single and double
  ionisation of helium by electron impact},
  \href{https://doi.org/10.1088/0953-4075/21/15/019}{\emph{Journal of Physics
  B: Atomic, Molecular and Optical Physics} {\bfseries 21} (1988) 2751}.

\end{thebibliography}\endgroup

\end{document}